\documentclass[aip,cha,reprint]{revtex4-1}
\usepackage{revsymb,graphicx,amssymb,amsmath,natbib,placeins,bm,color}

\usepackage{setspace}
\usepackage{verbatim}

\newcommand{\bi}{\begin{itemize}}
\newcommand{\ei}{\end{itemize}}
\newcommand{\ben}{\begin{enumerate}}
\newcommand{\een}{\end{enumerate}}

\newcommand{\bfi}{\begin{figure}[hbtp]}
\newcommand{\efi}{\end{figure}} 
\newcommand{\vepsi}{\varepsilon}
\renewcommand{\vec}{\mathbf}
\newcommand{\dr}{\partial}
\newcommand{\beq}{\begin{equation}}
\newcommand{\eeq}{\end{equation}}
\newcommand{\beqar}{\begin{eqnarray}}
\newcommand{\eeqar}{\end{eqnarray}}

\newcommand{\ba}{\begin{array}}
\newcommand{\ea}{\end{array}}
\newcommand{\St}{{\rm St}}

\begin{document}

\title{Levitation of heavy particles against gravity in asymptotically downward~flows}
   
\author{Jean-R\'egis Angilella}
\affiliation{Universit\'e de Caen Basse-Normandie, LUSAC, Cherbourg, France}

\author{Daniel J. Case}
\affiliation{Department of Physics and Astronomy, Northwestern University, Evanston, IL 60208, USA}

\author{Adilson E. Motter}
\affiliation{Department of Physics and Astronomy, Northwestern University, Evanston, IL 60208, USA}
\affiliation{Northwestern Institute on Complex Systems, Northwestern University, Evanston, IL 60208, USA}

\begin{abstract} 
In the fluid transport of particles, it is generally expected that heavy particles carried by a laminar fluid flow moving downward will also move downward. We establish a theory to show, however, that particles can be dynamically levitated and lifted by interacting vortices in such flows, thereby moving against gravity and the asymptotic direction of the flow, even when they are orders of magnitude denser than the fluid. The particle levitation is rigorously demonstrated for potential flows and supported by simulations for viscous flows. We suggest that this counterintuitive effect has potential implications for the air-transport of water droplets and  the lifting of sediments in water.
\end{abstract}

 \maketitle

\begin{quotation}
Levitation---the action of rising and hovering in apparent defiance of
gravity---is a fascinating phenomenon with many practical implications.
A classic demonstration is the  Bernoulli ball  levitation, in which a
macroscopic particle heavier than air (such as a ping pong  ball)
can 
levitate in response to an inclined upward air stream that appears to only partially balance gravity.
A key aspect of that form of levitation is the transversal stability due
to the Coand\v{a} effect,\cite{coanda}  which relies on the tendency
of the flow to curve around the surface of the ball and sustains
stable levitation when the upward air stream is tilted.
Here, we report a new form of fluid-dynamical levitation that can be
observed even for a downward stream (i.e., in the direction of
gravity)
and that allows heavy particles to be levitated by a flow
regardless of whether they are 10, 100, or 1,000 times denser than
the fluid. This phenomenon is fundamentally different from the Coand\v{a}
effect in that it concerns microscopic heavy particles and
requires no disturbance of the flow by the particles.
\end{quotation}

Many natural and  industrial flows transport small particles, like droplets,  sediments,  and  microorganisms.~\cite{review_advection} 
Inertial effects cause the trajectories of such particles to deviate from the streamlines of the flow, making the study of particle-laden flows challenging both theoretically and experimentally.~\cite{Cartwright2010,Volk2008,neutrally_buoyant, Pushkin2011} Key to such studies are the dissipative nature of the advection dynamics and the consequent tendency of the particles to accumulate in specific zones of the flow domain, both in closed flows~\cite{Maxey1987pof, McLaughlin1988, Squire02,Wang1992, Nishikawa2002, IJzermans2006, different_sizes_different_attractor_types, Angilella2010, inertial_particle_clustering,Ravichandran2014,inertial_particle_clustering,Nizkaya2010}  and in open flows,~\cite{inertial_open_flow_prl, inertial_open_flow_pre, Vilela2007, Haller2008, Angilella2014} even when the flows are incompressible. For example, particles less dense than the fluid tend to be attracted to the interior of vortices,~\cite{Annamalai1, Annamalai2} which can lead to the formation of attractors for the particle dynamics independently of the global properties of flow.  Particles denser than the fluid, on the other hand, tend to be repelled by vortices, which is a mechanism that
can lead to the formation of attractors if the flow is closed; 
this effect,  which is also related to the preferential trajectories phenomenon,~\cite{Squire02} has been widely investigated over the past two decades.~\cite{Cartwright2010}  However, much less is known about dense particles moving in flows that have unbounded streamlines and are therefore open. In open flows, an outstanding problem of particular interest concerns the transport of small particles much denser than the fluid, which we term {\it heavy particles} and which can represent for example water droplets in the~air.

In this 
article,
we 
 demonstrate
the possibility of levitation and upward transport of heavy particles by a flow moving asymptotically downward, 
even in the presence of gravity. 
Because at first  these conditions seem to facilitate downward 
advection, one might expect that all particles would  necessarily fall, which is in sharp contrast with the effect we report.
The starting point of our analysis is the observation that such a flow can support pairs of mutually interacting vortices traveling in a direction that opposes the flow. We thus focus on asymptotically simple flows  that move downward and have a pair of vortices moving upward. 
Using this class of flows we show that {\it attracting points} (dimension-zero attractors) formed near the center of vorticity can capture heavy particles released at any distance above the vortices. For this to occur,
particle inertia must allow for particles to approach the vortices and for the existence of attractors to retain them in that region; 
we show that these two conditions are satisfied for a wide range of Stokes  number in the class of flows we consider.   This is demonstrated analytically using asymptotic analysis and Melnikov functions, and is illustrated numerically using simulations in 
both
inviscid and viscous laminar~flows.
 
Our analysis is inspired by 
previous experimental realizations of flows with pairs of interacting vortices \cite{exp1,exp2,exp3,exp4,exp5} and
theoretical work on particle advection in such flows. \cite{Vilela2007, Angilella2010, Ravichandran2014,  Angilella2014, Nizkaya2010,Meleshko92,Pentek95,Govindarajan98,Muralidharan05}
Several studies have shown, 
both numerically \cite{Vilela2007} and analytically, \cite{Angilella2010, Ravichandran2014,  Angilella2014} the formation of attractors for the dynamics of heavy particles in the vicinity
of identical co-rotating vortices. 
Different work has shown that  such attractors persist for non-identical vortices in closed potential flows in the absence of gravity, \cite{Nizkaya2010} but no results exist for open flows,  viscous regimes, or the effects of gravity. Here, in order to demonstrate the proposed levitation of heavy particles, we first generalize the special results in open flows, previously established for  nongeneric vortex pairs, to the case of (i) generic pairs of both co-rotating and counter-rotating vortices of arbitrary vortex strength ratio,  
(ii)  for vortices moving against a background flow that is co-directional with gravity, and
(iii) for both potential and viscous flows.
Under these general conditions we then show the existence of attracting points 
that 
capture heavy particles from both the closed and open flow regions and  carry them against gravity and the background~flow.

   \begin{figure}[t!] 
        \centering
  \includegraphics[width=0.99\columnwidth]{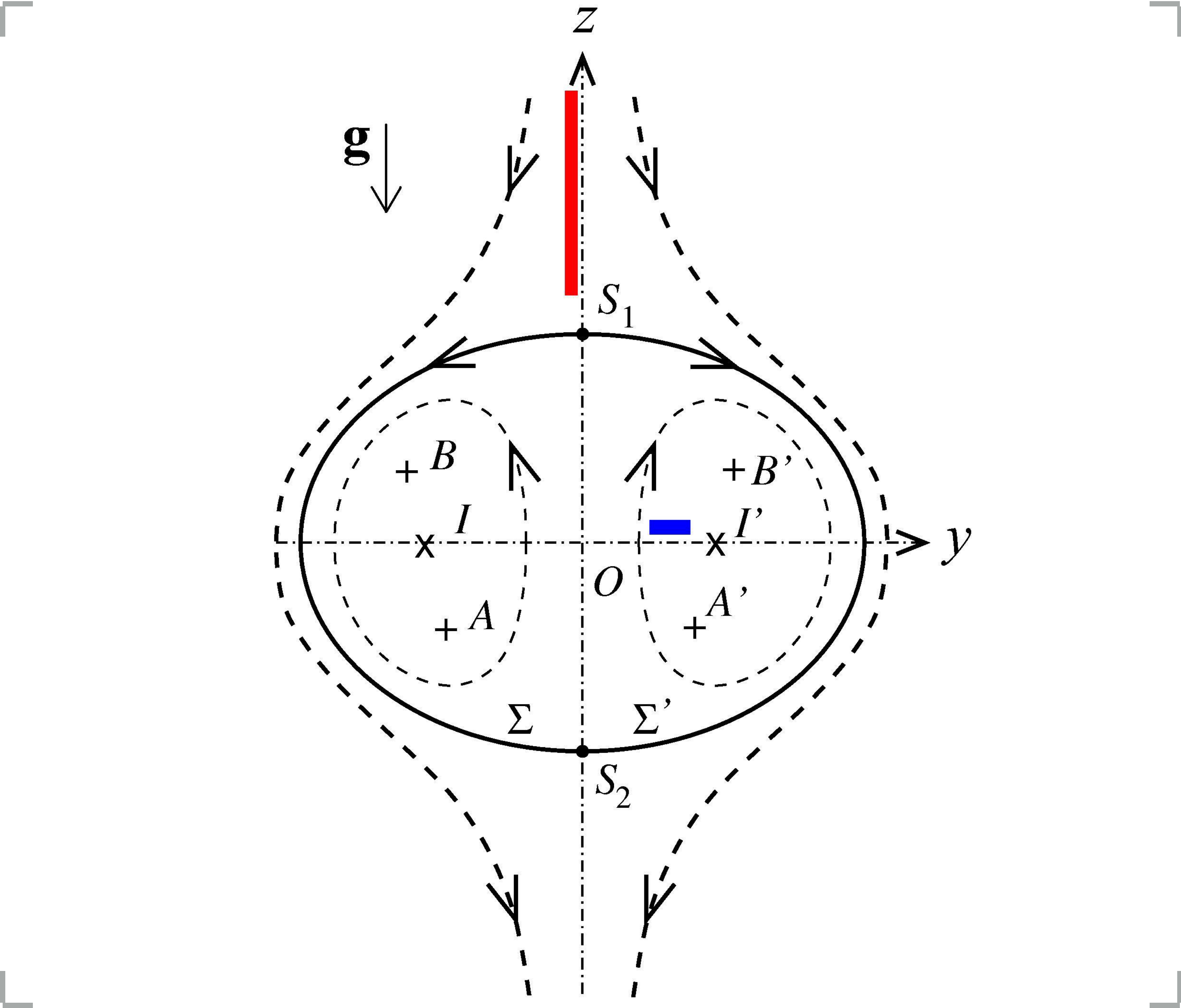}
          \vspace{-2mm}
\caption {Sketch of the flow: 
vortices $A$ and $B$ and their mirrors 
together with typical streamlines (dashed lines) in the reference frame 
of the centers of vorticity $I$ and $I'$. 
The flow along the open streamlines moves downward---the orientation of the gravitational field, $\vec g$.~Also shown is where the particles are initiated  
(with the velocity of the flow) 
in the potential flow simulations below: blue and red particles are released in the closed and open regions of the flow,~respectively.
           \vspace{-2mm} 
\label{sketch}}
\end{figure}

The model flow is
depicted in Fig.~\ref{sketch}. It consists of two vortices, $A$ and $B$, with strengths  $\Gamma_A$ and $\Gamma_B$, respectively, plus mirror vortices $A'$ and $B'$ with opposite strengths and symmetric positions with respect to the vertical axis $Oz$.  This axis can be regarded as a
 ``wall''   in the framework of potential flow theory used in our calculations.  The position $\vec {r}_I$ of the center of vorticity $I$ of the pair $(A, B)$  is 
 $
\vec{r}_I =  ({\Gamma_A \vec{r}_A + \Gamma_B \vec{r}_B})/({\Gamma_A+\Gamma_B}),
 $
where  $\vec r_A$ is the position  of  $A$ and $\vec r_B$ is the position of $B$.  
In the absence of mirror vortices, point $I$ remains fixed and  $AB$ rotates steadily 
around $I$ with angular velocity
$
\Omega_0 = ({\Gamma_A+\Gamma_B})/({8 \pi d_0^2}),
$
where $2 d_0 = |AB|$ is the   distance between the vortices, which remains constant in this case.
When the two pairs 
$(A,B)$ and $(A',B')$ 
are close enough to interact, elementary vortex dynamics show
that point 
 $I$  moves vertically, 
 and its distance to the symmetry axis $Oz$ remains equal to its initial value,
 $L_0$. 
For $d_0 \ll L_0$,
the (vertical) velocity of $I$ 
 approaches 
$
w_0 = ({\Gamma_A+\Gamma_B})/{4 \pi L_0}
$
in the  frame of the fluid at infinity.
The streamfunction of the exact 2D  potential
flow induced by the four vortices, in the  frame 
translating
at constant speed $w_0 \,\hat{\mathbf z}$,   
where $\hat{\mathbf z}$ is the 
upward unit vector, is  
\vspace{-2mm}
\beq
\psi(\vec r,t) = \frac{\Gamma_A}{4 \pi} \ln \frac{|\vec r-\vec {\bar r}_A |^2 }{|\vec r-\vec {r}_A |^2 } 
+ \frac{\Gamma_B}{4 \pi} \ln \frac{|\vec r-\vec {\bar r}_B |^2 }{|\vec r-\vec {r}_B |^2 }  + w_0 \, y,  
\label{exactPsi}
\eeq
where $\vec r=(y,z)$, and
$\vec {\bar r}_A$ and $\vec {\bar r}_B$ are
the positions  of the mirror  
vortices $A'$ and $B'$, respectively.  The 
instantaneous velocity of fluid elements in this frame trace
closed streamlines near the vortices ({\it closed flow} region) and open streamlines further away from them ({\it open flow} region). 
As indicated in Fig.~\ref{sketch}, these
two regions meet along the
separatrix streamline $\Sigma$ joining the
   stagnation points located on the $z$-axis, 
$S_1$ and $S_2$. 
  The structure formed by  $\Sigma \cup S_2 S_1$ is a heteroclinic cycle for the dynamics of fluid elements; 
heteroclinic (and homoclinic) cycles are generally expected to play a role 
in the transport of both  non-inertial and inertial particles.~\cite{ottino89,Cartwright2010} 

To proceed, we   define $\gamma = \Gamma_A/\Gamma_B$ as the vortex strength ratio, where $-1 < \gamma \le 1$. 
It can be checked that up to order 
$\vepsi^2$, when $\vepsi \equiv d_0/L_0 <  (\gamma+1)/2$,  
$A$ and $B$  rotate  around $I$ approximately as a rigid body with angular velocity  $\Omega_0$.
This allows us to write $\vec r_{A,B}(t)$ and ${\vec {\bar r}_{A,B}}(t)$ in the form of $2\pi/\Omega_0$-periodic functions plus $O(\vepsi^2)$ corrections. 
Throughout the rest of the 
article
we operate with the equations in non-dimensional form, 
using as units $L_0$  for lengths and $w_0/2$ for velocities 
since these choices capture the appropriate orders of magnitude
 for the flow near the heteroclinic cycle $\Sigma$. 
(No new notation is
introduced for non-dimensional variables.)
Expanding Eq.~(\ref{exactPsi}) in powers of $\vepsi$, in non-dimensional form  the streamfunction reads
\vspace{-1mm}
\begin{equation}
 \psi(\vec r,t) \!= \!  \psi_0(\vec r) + \frac{4 \gamma \vepsi^2 }{(1\!+\!\gamma)^2}\, 
\!\! \Big[ \psi_{2c}(\vec r)\cos \frac{2 t}{\vepsi^2} 
+  \psi_{2s}(\vec r)\sin \frac{2 t}{\vepsi^2} \Big] \!\!\! 
\label{psiasym}
\end{equation}
plus $O(\vepsi^3)$ terms, 
where  the  components $ \psi_0,\,\psi_{2c}, \, \psi_{2s} $ are 
identical to those
for the case $\gamma=1$ \cite{Angilella2011} given that the $\gamma$-dependence is accounted for by the prefactor (see 
supplementary material). 
The remainder in Eq.~(\ref{psiasym}) can be shown to be $O(\vepsi^4)$ when $\gamma=1$.
 The velocity 
field---defined  as $\vec u = \mathbf{\nabla}\times(\psi \,   \hat{\mathbf x})$, 
 where $\hat{\mathbf x}$ is the right handed unit vector orthogonal to the $yz$ plane---is therefore of the form 
$ \vec u(\vec r,t) = \vec u_0(\vec r) + \vepsi^2 \vec u_2(\vec r,t/\vepsi^2 ;\gamma) + O(\vepsi^3)$,  
where $\vec u_2$ is time-periodic  with period $T=\pi \vepsi^2$ and corresponds to the leading perturbation induced by the rotation of the vortices.  
   
Having established the fluid flow equations, we now write the
particle equation of motion in this flow. For 
a heavy 
 particle with small particle Reynolds number,  the non-dimensional equation is
$ \ddot{\mathbf r} = {\widetilde\St}^{-1}\left[{\mathbf u}(\mathbf r,t) + \widetilde{\mathbf V}_T  -  \dot{\mathbf r}\right]$,
where $\mathbf r$ is the particle position, $ \widetilde{\mathbf V}_T=- \widetilde{V}_T \hat{\mathbf z} = -2 (g\, \tau_p/w_0)\hat{\mathbf z}$ is the 
free-fall terminal velocity, and $\widetilde\St$ is the Stokes number (i.e.,  the response time of the particle, $\tau_p$, divided by the  time-scale of the flow, $2L_0/w_0$).~\cite{clift05}
 We also introduce another Stokes number, $\St = \Omega_0 \, \tau_p$, to describe the dynamics of particles directly influenced by the rotation of the vortices around each other. 
The formation of attractors near the vortices requires that  $\St$ 
be no larger than order one since drag has to balance centrifugal force in this case.  We therefore assume $\St = O(1)$, so that 
 $\widetilde\St \equiv \vepsi^2 \St\ll 1$ and the
 equation 
 for $\mathbf r$
can be reduced to 
\vspace{-1mm}
\beq
\dot{\mathbf r} = {\mathbf v}_0(\mathbf r) 
+ \vepsi^2 {\mathbf v}_2(\vec r,t/\vepsi^2;\gamma) + O(\vepsi^3), 
\label{rapidlyperturb}
\eeq
where 
$
\vec v_0(\vec r) = {\mathbf u}_0(\mathbf r) + \widetilde{\mathbf  V}_T  
$
and 
$
\vec v_2(\vec r,\tau;\gamma) \simeq  \vec u_2({\vec r},\tau;\gamma)  - \St \big[({\vec u}_0 + \widetilde{\vec V}_T) . \nabla {\vec u}_0 + {\dr {\vec u}_2  (\vec r,\tau;\gamma) }/{\dr \tau} \big],  
$
for $\tau=t/\vepsi^2$ 
(see also Ref.~\cite{Maxey1987jfm}). 
We show that the particle dynamics described by this equation have at least one (for $-1<\gamma\le 1$) and possibly two attracting points (for $0<\gamma\le 1$), provided that the Stokes number is not too large (see
supplementary material).

In Eq.~(\ref{rapidlyperturb}), the leading term,
$ {\mathbf v}_0$, represents the conservative dynamics of non-inertial particles in a steady flow induced by the equivalent to a  single vortex with strength $\Gamma_A+\Gamma_B$ (together with its mirror) plus a uniform flow $-\widetilde{V}_T  \hat{\mathbf z}$. The particle streamfunction for this term,  $\psi_p(y,z) = \psi_0(y,z) + y \widetilde{V}_T$,  is time-independent (a similar streamfunction has been used to describe plankton dynamics~\cite{Stommel1949}).  
The first perturbative term, $\vepsi^2 {\mathbf v}_2$, contains the contribution $\vec u_2$ of the unsteadiness of the flow due to the fact that for $\vepsi > 0$  there are two vortices rather than one, and also 
the effect of inertia in the $\St$ terms. 
It is thus convenient to regard this system as a time-independent Hamiltonian  $\psi_p$ perturbed by dissipative and fast periodic terms.~\cite{Gelfreich97}
To leading order in $\vepsi$, the trajectories of the particles coincide with the curves $\psi_p(y,z)=$ {\it cte}. These curves correspond to open streamlines separated from closed streamlines by a heteroclinic cycle, which we denote $\Sigma_p$ and which is the particle analog of $\Sigma$ in Fig.~\ref{sketch}  except that $\Sigma_p$ 
does not include the time-dependent perturbation terms.
 \begin{figure}[t!] 
        \centering
               \includegraphics[width=0.99\columnwidth]{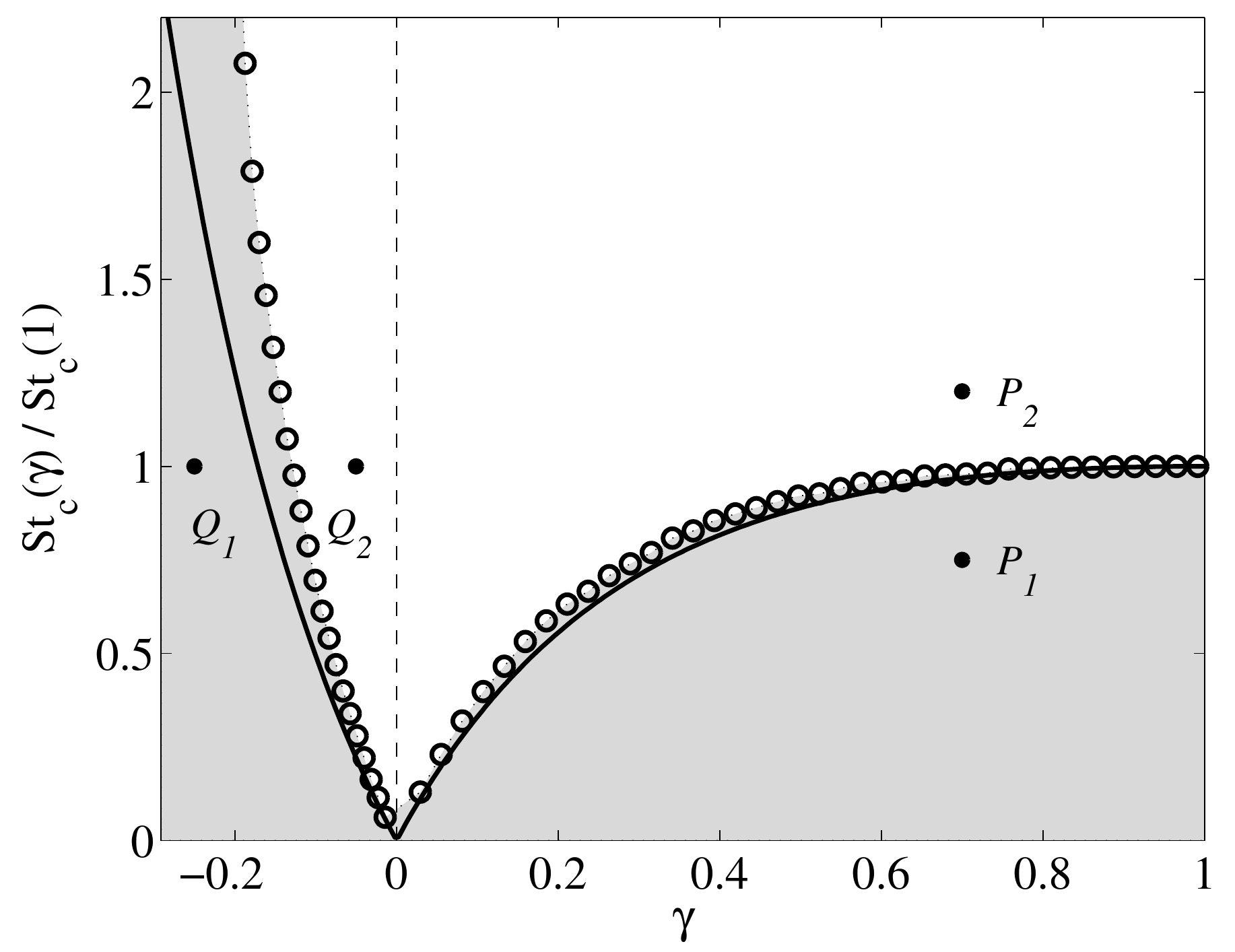}
          \vspace{-2mm}
\caption{Normalized critical Stokes number for arbitrary vortex  strength ratio, $\gamma$. Circles represent  
simulations
 for $\widetilde{V}_T=0.1$ and $\vepsi=0.35$,  whereas solid lines represent the theoretical prediction in Eq.~(\ref{Stc1}).  Particles released in the open flow 
 with $\St$
 in the shaded area 
 may cross $\Sigma_p$ and hence be levitated, while those above it~cannot.  
}
        \label{StcGamma}
           \vspace{-3mm}
\end{figure}

An important necessary condition for particles from the open flow 
to be captured by attracting points in the vicinity of the vortices is that they cross $\Sigma_p$ (the separatrix of the unperturbed dynamics) under the effect of the motion of the vortices.
The occurrence of separatrix crossing can be predicted 
employing a construction based on
separatrix  maps.~\cite{Chirikov1979,Kuznetsov1997} 
We consider a solution $\vec r(t)$ of the perturbed system (up to order $\vepsi^2$) and define $t_n \, (n =1,2,...)$ as the times at which the 
particle crosses the axis $Oy$ downward. 
We also use $\tau_{2n}$  and $\tau_{2n+1}$ to denote the times the particle passes closest to the saddle points $S_1$ and $S_2$, respectively,  and $H_n=\psi_p[\vec r(\tau_{2n})]$ and $H_{n+1}=\psi_p[\vec r(\tau_{2n+1})]$ to denote the corresponding values of the unperturbed Hamiltonian.
 Oscillations of $H_{n+1} - H_n$  around zero as $n$ varies will indicate that the separatrix $\Sigma_p$ 
is crossed.
Assuming that $\vec r(t) \simeq \vec q(t-t_n)$, for $\vec q(t)$ denoting the 
solution of the unperturbed system along $\Sigma_p$,
 we obtain
\begin{equation}
\vspace{-2mm}
H_{n+1} - H_n  \simeq \vepsi^2 M(t_n),
\label{sepmap1}
\end{equation}
where $M(t_n)$ is the Melnikov function associated with 
$\Sigma_p$:
\beq
\vspace{-2mm}
M(t_n) = \frac{4 \gamma   (a+ \widetilde{V}_T  b )}{(1+\gamma)^2} 
\Big[\sin \frac{2 t_n }{ \vepsi^2} - 2 \St \cos \frac{2 t_n }{ \vepsi^2}\Big]
 - m \, \St
\label{Melnikfn}
\eeq
(see 
supplementary material
for details). Here, the amplitudes $a=a(\vepsi)$ and $b=b(\vepsi)$ are functions of $\vepsi$ only, and 
$m$ is a constant accounting for the centrifugal effect along  $\Sigma_p$  due to the particle's inertia.

The Melnikov function  in Eq.~(\ref{Melnikfn}) is either strictly negative or oscillates between positive and negative values.~\cite{reference_comment} When the function has a constant negative sign,  we have $H_{n+1} < H_n$ for all $n$,  which  
indicates that particles are centrifuged away from the vortices. 
When the function has simple zeros, particles can enter and exit the closed flow. 
Thus, the  central prediction of our theory is that a heavy particle in the open flow 
may be captured by an attracting point near the vortices provided  that the Stokes number $\St$ is below the critical value  
\vspace{-3mm}
\beq
\St_c(\gamma) = \frac{4 \mid \gamma \mid   }{m (1+\gamma)^2} 
\Big[ a(\vepsi) + \widetilde{V}_T b(\vepsi) \Big],
\label{Stc}
\eeq
which is the condition for $M(t_n)$ to change sign (and in fact have infinitely many isolated zeros).
We therefore predict that the levitation of heavy particles released in the open 
flow region above the vortices is possible for $\St < \St_c(\gamma)$. These results imply that heavy particles with densities across many orders of magnitude can be levitated by the same flow.~\cite{comment}

 \begin{figure}[t!] 
        \centering
            \includegraphics[width=0.99\columnwidth]{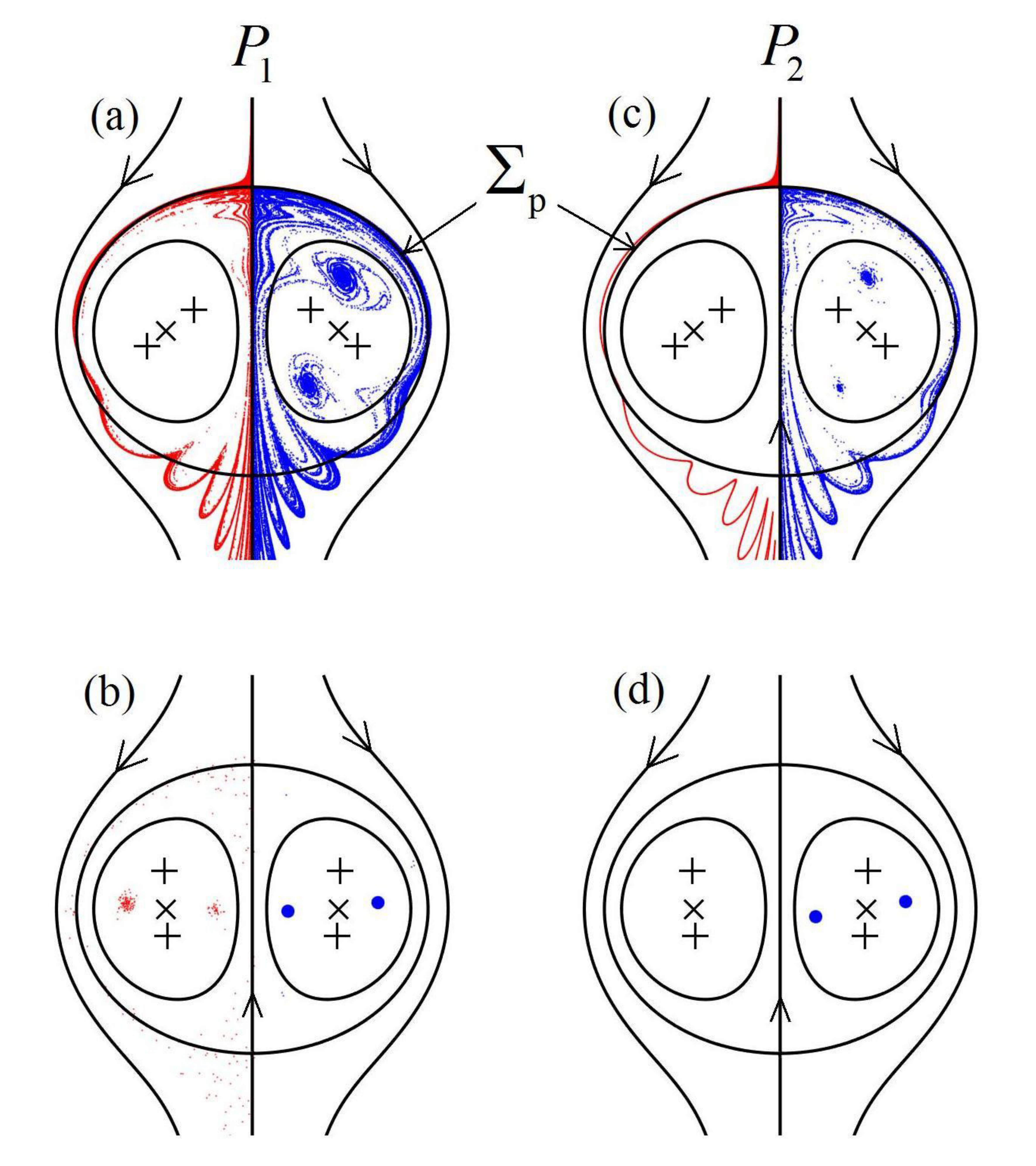}
                  \vspace{-2mm}
\caption{Levitation by co-rotating vortices
for the flow in Eq.~\ref{exactPsi}.
Colors indicate particle clouds at different times for (a, b) point $P_1$  ($\St <\St_c$)
and  (c, d) point $P_2$  ($\St >\St_c$)
in  Fig.~\ref{StcGamma}. 
The first row [(a) and (c)] corresponds to $t=16$ and second row [(b) and (d)]  to $t=95$  (in units of the undisturbed turnover time, $2\pi/\Omega_0$).
Also shown are streamlines defined by $\psi_{p}(y,z) = cte$ (continuous lines), the vortices ($+$), and  the center of vorticity ($\times$).  Red and blue represent particles released 
from the open and closed flows, 
respectively, and are shown 
on opposite sides of the wall to facilitate visualization.   
}
        \label{CloudsP1P2}
                   \vspace{-2mm}
\end{figure}

Remarkably, Eq.~(\ref{Stc}) shows that $\St_c$ is an increasing function of $\widetilde{V}_T$ and hence that gravity {\it facilitates} levitation. This means that a particle that
would be too inertial to penetrate inside $\Sigma_p$ in the absence of gravity can be captured by the attracting points 
when gravity is present.
This equation also shows that $\St_c$ depends on the vortex strength ratio $\gamma$ and in particular that the normalized critical Stokes number is
\vspace{-0.1cm}
\beq
\frac{\St_c(\gamma)}{\St_c(1)} = \frac{4 \mid \gamma \mid   }{(1+\gamma)^2},
\label{Stc1}
\eeq
irrespective of $\widetilde{V}_T$  and 
$\vepsi$. 
Thus, not only levitation is possible for both co- and counter-rotating vortices of arbitrary $\gamma\neq-1$, but also the phenomenon is {\it more pronounced} for counter-rotating vortices with sufficiently small $\gamma$  than for identical co-rotating vortices ($\gamma=1$). 

Figure~\ref{StcGamma} shows the analytical prediction in Eq.~(\ref{Stc1}) 
along with a numerical verification  for 
particles released
 in the open flow 
 above the vortices (red region in Fig.~\ref{sketch}), where bisection in $\St$ was used  to determine the critical value at which particles start crossing $\Sigma_p$. 
 The agreement  with the numerics is good, particularly for co-rotating vortices. 
Importantly, the numerical $\St_c$ is always higher than the 
theoretical prediction, widening the range of the effect.~\cite{comment_aggremment}

To further illustrate the theoretical results above we have done a series of computations by choosing the parameters according to the diagram of Fig.~\ref{StcGamma}. 
Figure~\ref{CloudsP1P2} shows the evolution of clouds of particles  for  two different Stokes numbers
in the case of co-rotating vortices:   
 $\St < \St_c$ [Fig.~\ref{CloudsP1P2}(a, b)] and  
 $\St > \St_c$ [Fig.~\ref{CloudsP1P2}(c, d)],  
 corresponding to the points $P_1$ and $P_2$ 
 in Fig.~\ref{StcGamma},  respectively.
The particles are  released in the open-streamline region above the vortices (red) and in the closed-streamline region near the vortices (blue), 
as sketched  in Fig.~\ref{sketch}.
The clouds are shown after $16$ turnover times [Fig.~\ref{CloudsP1P2}(a, c)] and after $95$ turnover times [Fig.~\ref{CloudsP1P2}(b, d)], which was chosen purposely large to facilitate visualization of the long-term behavior.  
For $P_1$, a fraction of  the blue particles
as well as a fraction of the red particles
accumulate near the two 
attracting points 
after long times. (For visualization,  particles near the attracting points were slightly dispersed). 
In contrast, for $P_2$ only blue particles are captured by the attracting points; red particles remain 
outside $\Sigma_p$ and are transported downstream.  
These observations are in complete agreement 
with our predictions. 

 \begin{figure}[t!]
        \centering
            \includegraphics[width=0.99\columnwidth]{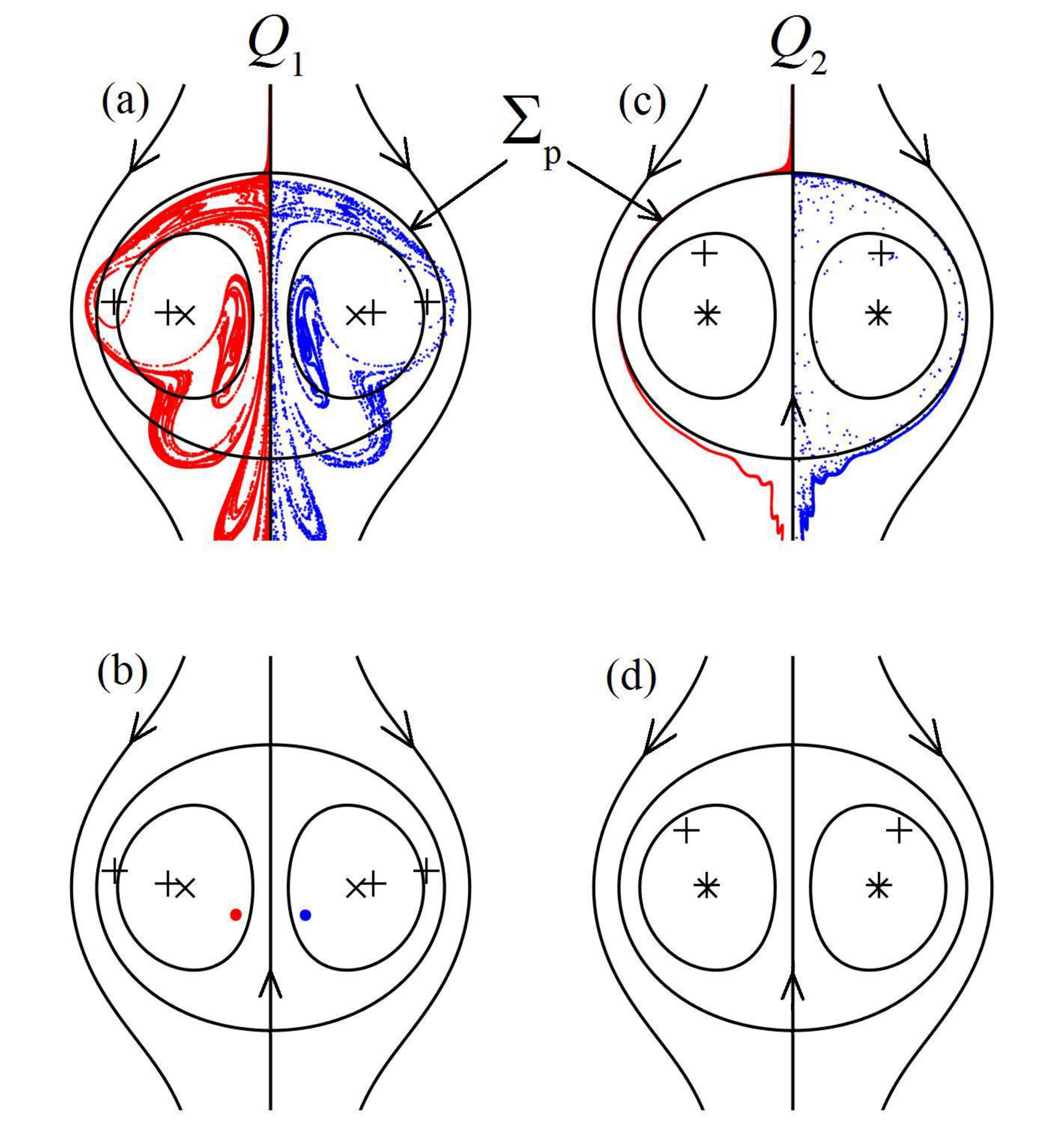}
                                 \vspace{-2mm}
\caption{Levitation by counter-rotating vortices. Counterpart of Fig.~\ref{CloudsP1P2} for (a, b) point $Q_1$ ($\St <\St_c$)
and  (c, d) point $Q_2$ ($\St >\St_c$)
in Fig.~\ref{StcGamma}. 
The first row [(a) and (c)] corresponds to $t=21$ and the second row [(b) and (d)] to $t=127$.
As in the case of co--rotating vortices, $\Sigma_p$ is open to particles released outside when $\St <\St_c$ and closed to those particles when $\St >\St_c$.  Here there is, however, a single attracting point for each vortex~pair for $Q_1$ and none for $Q_2$.
}
        \label{CloudsQ1Q2}
                   \vspace{-3mm}
\end{figure}

Similar agreement is observed for counter-rotating vortices,  
as shown in  
Fig.~\ref{CloudsQ1Q2} for
$\St < \St_c$  ($Q_1$ in Fig.~\ref{StcGamma}) and 
$\St > \St_c$ ($Q_2$ in Fig.~\ref{StcGamma}), 
but with two important differences. First, in the case of $Q_1$ there is only one attracting point around which a portion of blue and red particles accumulate. This is expected 
for the range of Stokes number considered, but 
studies of isolated counter-rotating pairs suggest that other attractors may exist for different parameters.~\cite{Nizkaya2010} Second, for $Q_2$, there is no attractor, so that not only do red particles not cross $\Sigma_p$ inward, but also all blue particles are centrifuged outward across $\Sigma_p$. As a result, in this case no particle is captured by an attractor independently of the initial condition.  
  
 \begin{figure}[t!]                         
        \centering
 \includegraphics[width=0.99\columnwidth]{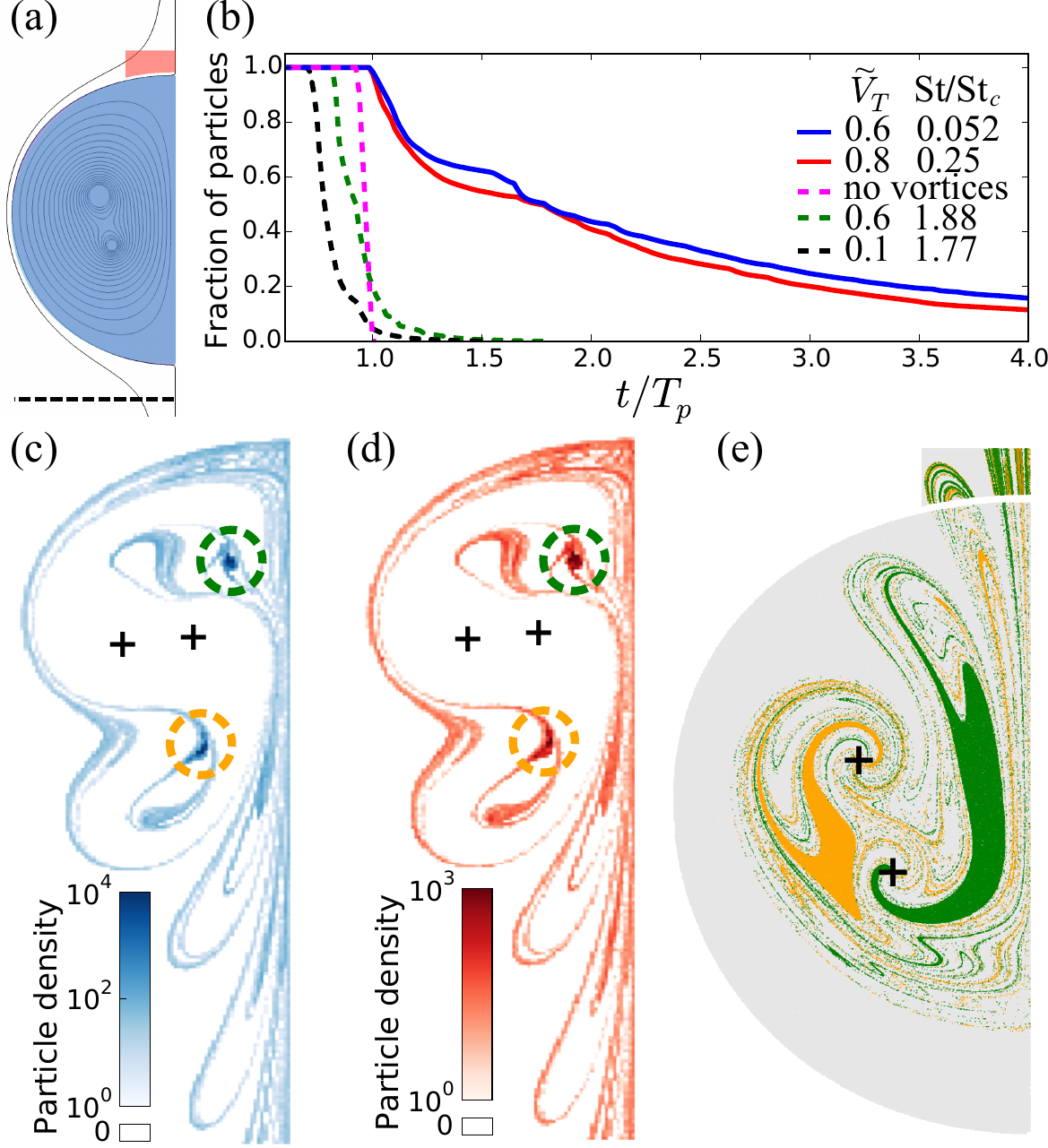}
                  \vspace{-2mm}
\caption{
Particle levitation for Navier-Stokes simulations of co-rotating vortices.
(a)~Initial positions of uniformly distributed particles with zero initial velocity in the open (red) and closed (blue) flow.
(b)~Fraction of particles from the open flow remaining above a threshold [dashed line in (a)] as a function of time in units
of $T_p$ (the maximum particle residence time  in the absence of vortices).
The various choices of $\widetilde{V}_T$ and $\St$ represent scenarios in which: particles are levitated by attracting points (solid lines),
particles do not cross into the closed flow region (dashed black),
no attracting point are present (dashed green), 
and vortices are absent (dashed magenta). 
(c, d) Density plots for particles from the closed (c) and open (d) flows at $t/T_p = 3.7$ for 
one choice of parameters in (b) (solid red);  dashed circles mark the positions of the attracting points.
(e) Section of the attraction basins: initial conditions of particles inside dashed circles of corresponding colors in (c) and (d).
The flow parameters are $\gamma = 0.7$, $\varepsilon = 0.375$, and Reynolds number $Re = 4,000$.
}
        \label{NS_fig}
                   \vspace{-2mm}
\end{figure}

To verify the significance of our predictions for more realistic flows, we have simulated viscous flow solutions of the Navier-Stokes equations  in the setup of Fig.~\ref{sketch}. 
As shown in Fig.~\ref{NS_fig}, for $\St < \St_c$,  particles with initial positions in both the closed flow  [\ref{NS_fig}(a, c)] and open flow [\ref{NS_fig}(a, d)]  are captured and levitated by  
attracting points in the vicinity of the vortices. The main difference from the case of  idealized potential flows considered
above is that levitation is not permanent in viscous flows since the vortices eventually coalesce. For the conditions considered in Fig.~\ref{NS_fig},
which corresponds to approximately 15 turnover times before vortex merging, particles from the open flow are accelerated downward by vortices
alone but, when capture occurs,
 up to $20\%$ of the particles are levitated by the vortices until they merge  [dashed versus continuous lines in Fig.~\ref{NS_fig}(b)].
A partial view of the basins of attraction [Fig.~~\ref{NS_fig}(e)] provides further insight into the initial conditions of the particles that can be levitated by 
this mechanism. Simulations were performed using OpenFOAM. \cite{OpenFOAM} For more details
and simulation movies, we refer to the supplementary material.

Our demonstration that heavy particles can be levitated shows that they can
be transported in any direction relative to the asymptotic flow and gravity.
Exploring  this  
 effect
in  more complex problems
(possibly involving non-laminar flows), 
such as 
 the air-transport of 
 water droplets and 
aerosols, \cite{ power_law_passive_advection,Iafrati2013,sea_salt_on_precipitation}
the resuspension 
of sediments by coherent vortical structures,~\cite{vertical_suspension_particle,bed_load_transport}
and industrial applications for particle sorting, \cite{inertial_open_flow_prl,inertial_open_flow_pre}
are among the questions of
great interest for future research.  

\section*{Supplementary Material}
The supplementary material includes the components of the streamfunction in Eq. (2), analysis of the attracting points, additional details on the Melnikov function calculation and  viscous flow simulations, Supplementary Figures for viscous flows, and Supplementary Movies showing animated versions of the dynamics in Fig. 5.

\begin{acknowledgments} 
This research was supported by NSF Grant  PHY-1001198.
\end{acknowledgments}

\widetext
\clearpage
\setcounter{equation}{0}
\setcounter{figure}{0}
\setcounter{table}{0}
\renewcommand{\theequation}{S\arabic{equation}}
\renewcommand{\thefigure}{S\arabic{figure}}
\renewcommand{\bibnumfmt}[1]{[S#1] }
\renewcommand{\citenumfont}[1]{S#1}





\centerline{\LARGE Supplementary Material}
\vspace{0.5cm}

\noindent
Levitation of heavy particles against gravity in asymptotically downward flows\\
Jean-R\'egis Angilella,  Daniel J. Case, and Adilson E. Motter \\




\bigskip
{\large\bf  Components of the Streamfunction}
\bigskip
\bigskip

The components $\psi_0$, $\psi_{2c}$, and  $\psi_{2s}$ of the steamfunction in Eq.~(2) of the main text read
\begin{eqnarray}
\psi_0(\mathbf{r}) &=&  \;\; \; \ln\frac{[z^2 + (y-1)^2]^2}{[z^2 + (y+1)^2]^2}  + 2y    , \\
\psi_{2c}(\mathbf{r}) &=&
-8y\,{\frac {\left [3\,{z}^{4}-\left (-1+{y}^{2}\right )^{2}+2\,{z}^{2
}\left (1+{y}^{2}\right )\right ]}{\left[{z}^{2}+\left (1+y\right )^
{2}\right ]^{2}\left[{z}^{2}+\left (1-y\right )^{2}\right]^{2}}} , \\
\psi_{2s}(\mathbf{r}) &=& 
-8yz\,{\frac {\left[-3+{z}^{4}+2\,{y}^{2}+{y}^{4}+2\,{z}^{2}\left (-1
+{y}^{2}\right )\right]}{\left[{z}^{2}+\left (1+y\right )^{2}
\right]^{2}\left[{z}^{2}+\left (1-y\right )^{2}\right ]^{2}}},
\end{eqnarray}
where $\mathbf{r}=(y,z)$ (as indicated in Fig.~1).

\bigskip
\bigskip
\medskip
{\large\bf Dynamics Near the Attracting Points} 
\bigskip

\baselineskip 17.5pt

\bigskip
To prove the existence and stability of attracting points when gravity is present,  we focus on one vortex pair,  say $(A,B)$. The pair $(A',B')$ is assumed to be far  
from $(A,B)$ so that, to leading order, the angular velocity is
$\Omega_0$ and the translational velocity of point $I$ is $w_0$. We consider the motion of heavy particles in the reference frame rotating at constant velocity $\Omega_0$, and translating at constant vertical speed $w_0$. We also non-dimensionalize the variables using time and length scales relevant for the dynamics near the vortices: $1/\Omega_0$ for times and $d_0$ for lengths. These units will be called ``internal units" in the following and should not be confused with the ``external units" $2L_0/w_0$ and $L_0$ used in the main text  
to investigate the dynamics near the separatrix $\Sigma_p$.
In addition, we use the rotating Cartesian coordinates $(I;X,Y,Z)$ where $IX=Ix$ is perpendicular to the
plane of the flow, and $IY$ and $IZ$ coincide with $Iy$ and $Iz$ at the initial time.
Using $\hat{\mathbf X}$, $\hat{\mathbf Y}$, and $\hat{\mathbf Z}$ to denote the unit vectors in the coordinate directions,
the unit vector in the direction of gravity 
reads $\hat{\mathbf g}=-\hat{\mathbf Y} \sin t  -  \hat{\mathbf Z} \cos t $.

In this new frame, the non-dimensional equation of motion for heavy particles in the internal system of units is
\begin{equation}
\ddot{\mathbf{R}} = \frac{1}{\mbox{St}}\big(\mathbf{W} - \dot{\mathbf{R}}  \big) + \frac{V_T}{\mbox{St}} \hat{\mathbf g} + \mathbf{R} - 2 \hat{\mathbf X} \times \dot{\mathbf{R}},
\label{dynrotframe}
\end{equation}
where ${\mathbf{R}}$ is the particle position and $\mathbf{W}$ is the fluid velocity field (in this frame). The last two terms are the centrifugal and Coriolis forces, respectively.
Because particles are much heavier than the fluid, forces proportional to the mass of the displaced fluid, like the Archimedes force and the opposite of the Coriolis and centrifugal forces acting on the fluid, have been neglected.
Also, St=$\Omega_0 \tau_p$ is the Stokes number already introduced in the main text, $V_T = g \tau_p/\Omega_0 d_0$ is the non-dimensional settling velocity in still fluid, and the ratio $V_T/$St is the inverse Froude number. The Stokes number (St) and non-dimensional settling velocity ($V_T$) introduced with the internal units are related to the  external ones by
\begin{equation}
\widetilde {\mbox{St}} = \vepsi^2 \mbox{St},\quad \mbox{and} \quad \widetilde V_T = V_T/\vepsi,
\end{equation}
where $\widetilde {\mbox{St}}\ll 1$, $\mbox{St} = O(1)$, $\widetilde V_T = O(1)$, and $V_T\ll 1$.  
Therefore, gravitational settling is weak in the internal dynamics  (near the vortices) and stronger in the external one (near the separatrix $\Sigma_p$).
In contrast, particle inertia effects are weak in the  external dynamics (which favors the crossing of $\Sigma_p$) and stronger near the vortices (which favors capture by attracting points).  

 To order $\vepsi^2$, the fluid velocity field is
\begin{equation}
\mathbf{W} = \mathbf{W}_0(\mathbf{R},\gamma) + \vepsi^2 \big[ \mathbf{W}_{2c}(\mathbf{R},\gamma) \cos 2 t + \mathbf{W}_{2s}(\mathbf{R},\gamma) \sin 2 t  \big],
\end{equation}
where the leading order is the relative velocity field induced by an isolated vortex pair with
vortex strength ratio $\gamma$ (see, e.g., Ref.~\cite{sNizkaya2010}) and the velocity fields
$\mathbf{W}_{2c}$ and $\mathbf{W}_{2s}$ account for 
the effect of the vertical wall (the symmetry axis $Oz$ separating $AB$ and $A'B'$). 
In the limit of small $\vepsi$ the wall causes each vortex pair  to stretch and compress
twice each revolution. That is why, in this frame and system of units, the perturbation terms are proportional to $\cos 2 t$ and $\sin 2t $.


To analyze the stability of equilibrium points, we follow the procedure employed by IJzermans and Hagmeijer in Ref.~\cite{sIJzermans2006}, and set
 \begin{equation}
 \mathbf{R}(t) = \mathbf{R}_{eq} + \mathbf{h}(t),
 \label{decomp}
 \end{equation}
 where $\mathbf{R}_{eq}$ denotes any one of the equilibrium 
points in the rotating frame in the absence of both wall and gravity 
(the existence of these points, when neither wall nor gravity is present,  has been proven in Ref.~\cite{sAngilella2010} for identical vortices and in Ref.~\cite{sNizkaya2010} 
 for unequal vortices). 
In particular,  $\mathbf{R}_{eq}$ is any solution of $\mathbf{W}_0(\mathbf{R}_{eq})/\mbox{St} + \mathbf{R}_{eq} = \mathbf{0}$, which reflects the balance between inward drag and the centrifugal force at the equilibrium 
point. In Eq.~(\ref{decomp}), $\mathbf{h}=\mathbf{h}(t)$ represents the perturbation accounting for the effect of gravity and the wall.
Using this decomposition in Eq.~(\ref{dynrotframe}), and neglecting the quadratic terms in 
$\mathbf{h}$, we obtain
\begin{eqnarray}
 \ddot{\mathbf{h}} = \frac{1}{\mbox{St}}\Big\{({\text D}\mathbf{W}_{0}^{eq})\, \mathbf{h}
+ \vepsi^2 \Big[\mathbf{W}_{2c}(\mathbf{R}_{eq},\gamma) \cos 2 t + \mathbf{W}_{2s}(\mathbf{R}_{eq},\gamma) \sin 2 t \nonumber \\
- \widetilde V_T(\hat{\mathbf Y} \sin t  +  \hat{\mathbf Z} \cos t) \Big] - \dot{\mathbf{h}} \Big\}
 + {\mathbf{h}} - 2 \hat{\mathbf X} \times \dot{\mathbf{h}}\, ,
\end{eqnarray}
where ${\text D}{\mathbf f}$ is used to denote the Jacobian matrix of ${\mathbf f}$ and ${\text D}\mathbf{W}_{0}^{eq} ={\text D}\mathbf{W}_{0}|_{\mathbf{R}_{eq}}$.
The general solution of this non-homogeneous linear equation is the sum of a particular solution $\mathbf{h}_a(t)$ and the solution $\mathbf{h}_b(t)$ of the homogeneous part of the equation.  

The homogeneous equation corresponds exactly to the case in which both gravity and the wall are absent. 
Focusing on equilibrium points $\mathbf{R}_{eq}$ that (in the absence of gravity and the wall) are asymptotically stable, we have
 $\mathbf{h}_b(t) \to 0$ as $t \to +\infty$.  
Provided the internal Stokes number is not too large, as considered here,  there are either one or two such stable points when neither gravity nor the wall is present~\cite{sNizkaya2010};  from here on we consider only these equilibria and show that they are converted into time-dependent attracting points when the effect of gravity and/or the wall are significant.

The particular solution  can be sought in the form of a combination of two Fourier modes:
\begin{equation}
\mathbf{h}_a(t)= \sum_{n=1,2}(\mathbf{p}_n \cos n t + \mathbf{q}_n \sin n t ),
\label{ha}
\end{equation} 
where the amplitudes $\mathbf{p}_n$ and $\mathbf{q}_n$ are given by a set of linear equations,
\begin{eqnarray}
\mathbf{L}_1 \mathbf p_1 - \mathbf{M}_1 \mathbf q_1 &=& \vepsi^2 \frac{\widetilde V_T}{\mbox{St}} \hat{\mathbf Z},\\
\mathbf{L}_1 \mathbf q_1 + \mathbf{M}_1 \mathbf p_1 &=& \vepsi^2 \frac{\widetilde V_T}{\mbox{St}} \hat{\mathbf Y},\\
\mathbf{L}_2 \mathbf p_2 - \mathbf{M}_2 \mathbf q_2 &=& - \frac{\vepsi^2}{\mbox{St}}   \mathbf{W}_{2c}(\mathbf{R}_{eq},\gamma),\\
\mathbf{L}_2 \mathbf q_2 + \mathbf{M}_2 \mathbf p_2 &=& - \frac{\vepsi^2}{\mbox{St}}   \mathbf{W}_{2s}(\mathbf{R}_{eq},\gamma),
\end{eqnarray}
for $\mathbf{L}_1 = \frac{\mbox{1}}{\mbox{St}} {\text D}\mathbf{W}_{0}^{eq} + 2\mathbf{I} $,
 $\mathbf{M}_1 = \frac{\mbox{1}}{\mbox{St}} \mathbf{I} + 2 \mathbf{A}$, $\mathbf{L}_2 = \mathbf{L}_1+3\mathbf{I}$, $\mathbf{M}_2=2\mathbf{M}_1$, and matrix $\mathbf{A}$ representing a rotation of $\pi/2$ around $\hat{\mathbf X}$. 
The matrices $\mathbf{M}_1$ and $\mathbf{M}_2$ are invertible---it can be checked that their determinants are nonzero irrespective of $\mbox{St}$.
The matrix $\mathbf{L}_1$ is  also invertible since its eigenvalues are $\lambda_i=2 + \mu_i/\mbox{St}$ ($i=1,2$), where $\mu_i$'s are the eigenvalues of the matrix
 ${\text D}\mathbf{W}_{0}^{eq}$, which are known to have non-zero imaginary parts. 
 Finally, a similar argument can be used to show that the matrix $\mathbf{L}_2$ is also invertible. Thus, after some elementary algebra, we are led to
\begin{equation}
\big(\mathbf{M}_1^{-1} \mathbf{L}_1 + \mathbf{L}_1^{-1} \mathbf{M}_1\big) \mathbf p_1
= \vepsi^2 \frac{\widetilde V_T}{\mbox{St}} \big(  \mathbf{L}_1^{-1}\hat{\mathbf Y}+\mathbf{M}_1^{-1}\hat{\mathbf Z}\big).
\end{equation}
By solving this last system we obtain the amplitude $\mathbf p_1$. The other amplitudes $\mathbf p_n$ and $\mathbf q_n$  can be obtained in a similar way,
so that the periodic solution $\mathbf{h}_a(t)$ in Eq.~(\ref{ha})  exists and is uniquely defined.

We therefore conclude that, for sufficiently small $\vepsi \ll 1$, the   
trajectories of inertial particles in the presence of gravity  and the wall converge to some periodic orbit in an $\vepsi^2$-neighborhood of the equilibrium point $\mathbf{R}_{eq}$ of the unperturbed system. 

\bigskip
\bigskip
\medskip
{\large\bf  Calculation of the Melnikov Function}
\bigskip
\bigskip

The variation of the undisturbed Hamiltonian $H=\psi_p(\mathbf r(t))$ along the disturbed trajectory $\mathbf r(t)$ between the discrete times $\tau_{2n}$ and $\tau_{2n+1}$ 
is~\cite{sChirikov1979,sKuznetsov1997}
\begin{equation}
 H_{n+1}-H_n = \int_{\tau_{2n}}^{\tau_{2n+1}} \frac{d}{dt}\psi_p(\mathbf r(t))\, dt
 = \int_{\tau_{2n}}^{\tau_{2n+1}} \nabla \psi_p ({\mathbf r} (t))   \cdot \frac{d{\mathbf  r} (t)}{dt}\, dt .
\end{equation}
 Assuming the trajectory is close to the separatrix $\Sigma_p$, we write $\mathbf r(t) \simeq \mathbf q(t-t_n)$, where $\mathbf q(t)$ is a trajectory on $\Sigma_p$. Therefore it satisfies $\dot {\mathbf q} = \mathbf v_0(\mathbf q)$,   $\mathbf q(-\infty)=S_1$, and $\mathbf q(+\infty)=S_2$, where the $z$-coordinates of $S_1$ and $S_2$ are given by $z=\pm \big[(6-\widetilde{V}_T)/(2+\widetilde{V}_T)\big]^{1/2}$. For convenience, we take the initial condition $\mathbf q(0)$ at the intersection between $\Sigma_p$ and the axis $Oy$.
In addition, we perform the change of variables $t \to t-t_n$, and take into account 
the fact that $\tau_{2n}-t_n < 0   < \tau_{2n+1}-t_n$ and that $|\tau_{2n}-t_n|\gg 1$ and $|\tau_{2n+1}-t_n|\gg 1$  (because the dynamics is very slow near $S_1$ and $S_2$). The integration interval $[\tau_{2n}-t_n , \tau_{2n+1}-t_n]$ can then be replaced by $[-\infty,+\infty]$, leading
to $H_{n+1}-H_n = \vepsi^2 M(t_n)$, where $M(t_n)$ is the Melnikov function,
 \begin{equation}
M(t_n) = 
 \int_{-\infty}^{\infty} \negmedspace \negmedspace \negmedspace \negmedspace 
{\mathbf v}_0({\mathbf q}(t)) \times {\mathbf v}_2\Big(\mathbf q(t),\frac{t+t_n}{\vepsi^2} \Big) \, dt.
 \end{equation}
Note that in this equation and others below,  the cross products should be interpreted as projected in the $ \hat{\mathbf x}$ direction.

The unperturbed trajectory ${\mathbf  q}(t)$ cannot be obtained analytically for 
{arbitrary}
$\widetilde{V}_T$. We  therefore approximate it by assuming that $\widetilde{V}_T$, though larger than
the internal non-dimensional velocity $V_T$, is sufficiently smaller than unity that it allows us to write
${\mathbf  q}(t) = {\mathbf q}_0(t) + \widetilde{V}_T\, {\mathbf  q}_1(t) + O(\widetilde{V}_T^2)$, where the functions  
${\mathbf  q}_0(t)$ and ${\mathbf  q}_1(t)$ are determined numerically.
 Setting $\mathbf U_2= \mathbf u_2 - \mbox{St} \, \partial \mathbf u_2/\partial \tau$,
 after some algebra 
  the Melnikov function reads
\begin{equation}
M(t_n) = I_1+I_2+I_3 + O( \widetilde V_T^2)+O( \widetilde V_T \mbox{St}),
\end{equation}
where
\begin{eqnarray}
I_1= \int_{-\infty}^{\infty} \negmedspace \negmedspace \negmedspace \negmedspace 
\dot{\mathbf  q}_0(t)  \times {\mathbf U}_2\Big(\mathbf q_0(t),\frac{t+t_n}{\vepsi^2}\Big)  \, dt 
 - \mbox{St}\int_{-\infty}^{\infty} \negmedspace \negmedspace \negmedspace \negmedspace \dot{\mathbf q}_0(t)  \times \ddot{\mathbf q}_0(t)  \, dt,\\
I_2= 
\widetilde V_T
 \int_{-\infty}^{\infty} \negmedspace \negmedspace \negmedspace \negmedspace 
\dot{\mathbf  q}_1(t) \times   \mathbf{U}_2\Big(\mathbf q_0,\frac{t+t_n}{\vepsi^2}\Big) \, dt,   \\
I_3= \widetilde V_T
 \int_{-\infty}^{\infty} \negmedspace \negmedspace \negmedspace \negmedspace 
\dot{\mathbf q}_0(t) \times \Big(({\text D} \mathbf{U}_2)\,   \mathbf  q_1(t) \Big) \, dt .  
\end{eqnarray}
By expanding the sines and cosines appearing in $I_1$ and assigning zero to the integrals of odd functions of $t$, we obtain
\begin{equation}
I_1 = \frac{4 \gamma}{(1+\gamma)^2} a(\vepsi)\Big(\sin \frac{2t_n}{\vepsi^2}-2 \mbox{St} \cos \frac{2t_n}{\vepsi^2}\Big)
 - m \, \mbox{St},
\end{equation}
where  the multiplicative coefficient $a(\vepsi)$ depends only on $\vepsi$,
\begin{equation}
a(\vepsi)=
\int_{-\infty}^{\infty} \negmedspace \negmedspace \negmedspace \negmedspace 
\dot{\mathbf  q}_0(t)  \times \Big[{\mathbf u}_{2s}(\mathbf q_0(t)) \cos \frac{2t}{\vepsi^2}-  
{\mathbf u}_{2c}(\mathbf q_0(t)) \sin \frac{2t}{\vepsi^2} 
 \Big]  \, dt, 
\end{equation}
 and  $m$ is a constant,
\begin{equation}
m= \int_{-\infty}^{\infty} \negmedspace \negmedspace \negmedspace \negmedspace  \dot{\mathbf  q}_0(t)  \times  \ddot{\mathbf  q}_0(t)  \, dt.
\end{equation}
The velocity fields $\mathbf u_{2c}$ and  $\mathbf u_{2s}$ correspond to the streamfunctions $\psi_{2c}$ and $\psi_{2s}$ respectively.

The constant $m$ is computed numerically once  $\mathbf q_0(t)$ has been determined, resulting in $m\simeq 30.4$. The function $a(\vepsi)$ is also computed numerically, and then fitted with an exponential-rational function of the form
\begin{equation}
a(\vepsi) = \frac{e^{-\alpha_e/\vepsi^2}}{\vepsi^2} \big(\alpha_0 + \alpha_2 \vepsi^2+ \alpha_4 \vepsi^4\big),
\end{equation}
where the $\alpha_i$'s are obtained using a least-square algorithm. The resulting constants are $\alpha_e \simeq 0.60$, $\alpha_0 \simeq 16.5$, $\alpha_2 \simeq -9.5$, and $\alpha_4 \simeq -51.5$. Applying the same treatment to the integrals $I_2$ and $I_3$, we obtain
\begin{equation}
I_2+I_3 = \frac{4 \gamma}{(1+\gamma)^2}\, \widetilde V_T  \big[b_2(\vepsi)+b_3(\vepsi)\big] \Big(\sin \frac{2t_n}{\vepsi^2}-2 \mbox{St} \cos \frac{2t_n}{\vepsi^2}\Big),
\end{equation}
where
\begin{eqnarray}
b_2(\vepsi)=
\int_{-\infty}^{\infty} \negmedspace \negmedspace \negmedspace \negmedspace 
\dot{\mathbf q}_1(t)  \times \Big[{\mathbf u}_{2s}(\mathbf q_0(t)) \cos \frac{2t}{\vepsi^2}-  
{\mathbf u}_{2c}(\mathbf q_0(t)) \sin \frac{2t}{\vepsi^2}  \Big]  \, dt ,\\
b_3(\vepsi)=
\int_{-\infty}^{\infty} \negmedspace \negmedspace \negmedspace \negmedspace 
\dot{\mathbf  q}_0(t)  \times \Big[({\text D}{\mathbf u}_{2s}) \, 
 \mathbf q_1(t) \, \cos \frac{2t}{\vepsi^2}-  ({\text D} {\mathbf u}_{2c}) \, 
\mathbf q_1(t) \,\sin \frac{2t}{\vepsi^2}
 \Big]  \, dt.
\end{eqnarray}
Both $b_2$ and $b_3$ depend only on $\vepsi$ and are computed numerically once ${\mathbf  q}_0(t)$ and ${\mathbf  q}_1(t)$  have been determined. The sum $b_2(\vepsi)+b_3(\vepsi)$, denoted $b(\vepsi)$ in the main text, is fitted with an exponential-rational function of same form as the one used for $a(\vepsi)$, with the coefficients $\alpha_i$'s replaced  by $\beta_i$'s: 
$\beta_e \simeq 0.60$, $\beta_0 \simeq 174.6$, $\beta_2 \simeq - 1120.6$, and $\beta_4 \simeq 1972.0$.

\bigskip
\bigskip
\medskip
{\large\bf  Navier-Stokes Simulations of Viscous Flows}
\bigskip
\bigskip

The computations in Figs.~2-4 (main text) 
were performed assuming idealized inviscid flows (exact solutions of the Euler equations). For moderate Reynolds number, viscosity is expected to have significant effects on the flow, as viscous diffusion causes the size of the vortex cores to expand over time. Once the radius of either core becomes comparable to the distance between the vortices, vortex merging occurs and any attractors of the particle dynamics vanish. We performed simulations solving the two-dimensional Navier-Stokes equations with a viscous working fluid to determine whether heavy-particle levitation could be observed prior to the eventual coalescence of the vortices. The initial velocity field of the fluid consisted of the superposition of a uniform downward flow (in the direction of gravity) with a velocity magnitude of $0.625 w_0$ relative to the laboratory reference frame and two pairs of co-rotating Lamb-Oseen vortices with inter-vortex distance $2d_0$ as in the setup of Fig.~1 (main text).  The Reynolds number of the flow, $Re = \Omega_0 {d_0}^2/\nu$, where $\nu$ is the dynamic viscosity of the fluid, was set to be $4,000$ in all simulations we report. 
For these conditions, in each run of our simulations we  released $100,000$ particles from the closed flow  and $50,000$  from the open flow  with zero initial velocity, as defined in Fig.~5(a) (main text). After simulating many combinations of $\widetilde{V}_T$ and $\widetilde {\mbox{St}} $, we observed the following scenarios:
for ${\mbox{St}}/{\mbox{St}}_c>1$ either no attracting points exist or they exist but no particles from the open flow are captured by them; 
for ${\mbox{St}}/{\mbox{St}}_c<1$ attracting points exist and particles from the open flow are captured and levitated by them. 

 \begin{figure}[t!]  
        \centering
               \includegraphics[width=.7\textwidth]{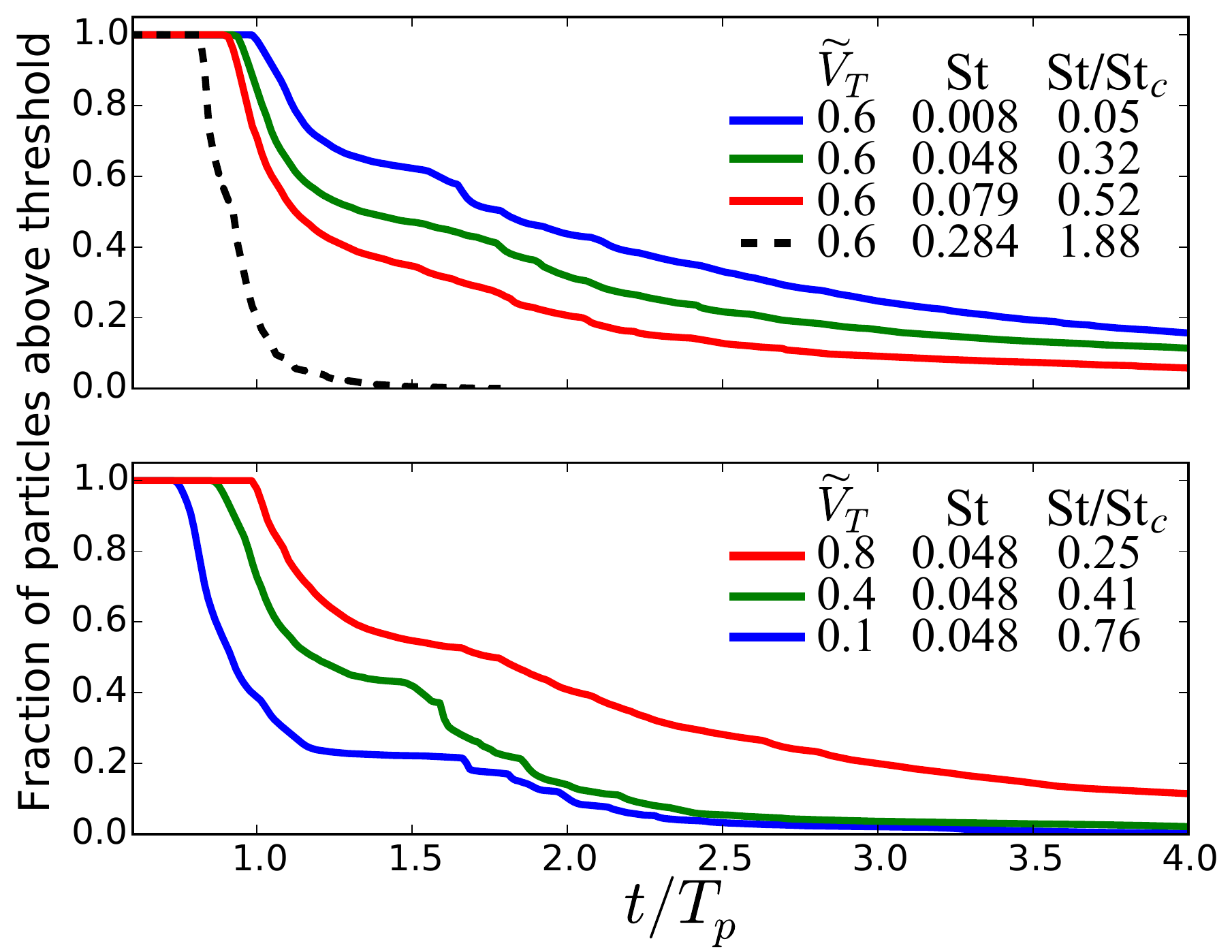}
                                 \vspace{-2mm}
\caption{\baselineskip 15.5pt  
Fraction of particles from the open flow that remain above the threshold marked in Fig.~5(a) in the main
text as function of time, for  $\gamma = 0.7$ and $\varepsilon = 0.375$.
The different curves correspond to 
(a) different choices of ${\mbox{St}}$ for $\widetilde{V}_T = 0.6$ and 
(b) different choices of $\widetilde{V}_T$ for ${\mbox{St}}  = 0.048$.    
The fraction of particles levitated for a long period increases  as ${\mbox{St}}$ is decreased  and
as $\widetilde{V}_T$ is (and hence gravity) increased, which parallels the dependence of $\mbox{St}_c$ in our theory based on potential flows.
}
        \label{figS1}
\end{figure}

To quantify the levitation mechanism for particles from the open flow, we calculate the fraction of particles released in the open flow 
that remain above a threshold as a function of time. We set the threshold to a distance $2.4L_0$ downstream from the initial center of vorticity $I$, which lies right below the closed flow separatrix so that any particle that falls below this threshold will not be pulled back upstream. This fraction is shown in Fig.~\ref{figS1}, illustrating the different scenarios mentioned above, where time is represented in units of the maximum time $T_p$ a particle with the given initial conditions would remain above threshold in the the same flow without vortices. We define the levitation period  to be the time  $t -T_p$ the particle remains above this threshold.
For example, for $\widetilde{V}_T = 0.6$ and ${\mbox{St}}/{\mbox{St}}_c  = 0.05$ particles from the open flow approach the attracting points and 
nearly $20\%$ of the particles have a levitation period larger than $3T_p$. For $\widetilde{V}_T = 0.6$ and ${\mbox{St}}/{\mbox{St}}_c  = 1.88$, on the other hand, there are no attractors in the vicinity of the vortices and the levitation period is close to zero for most particles. 
We also observe that for a fixed value of  ${\mbox{St}}$, increasing $\widetilde{V}_T $ can increase the fraction of particles above the threshold at all times, which is counter-intuitive but consistent with our analytical results for potential flow, where stronger gravity is observed to enhance the levitation effect.

Simulations were performed with OpenFOAM  (version $2.2$) using the PISO algorithm in solving for the flow field. \cite{spiso} The particle motion was  calculated with Lagrangian particle tracking with one-way coupling of the particles to the fluid.

\bigskip
\bigskip
\medskip
{\large\bf  Animated Visualization of Navier-Stokes Simulations}
\bigskip
\bigskip

\begin{figure}[h!]  
        \centering
               \includegraphics[width=.68\textwidth]{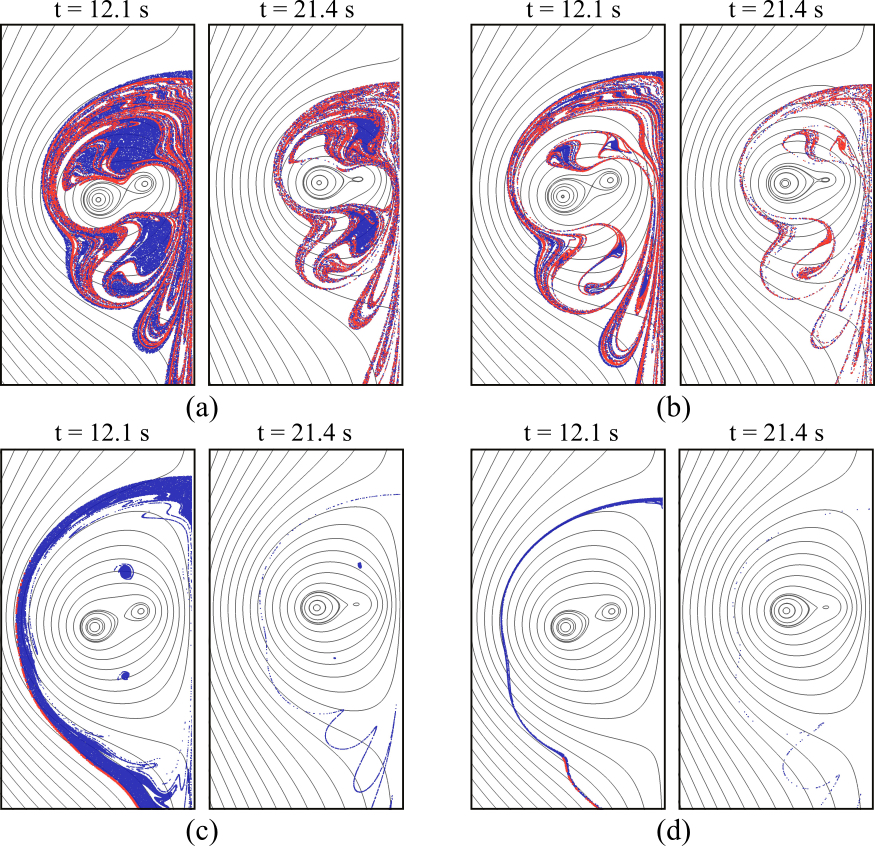}
                                 \vspace{-2mm}
\caption{\baselineskip 15.5pt 
Early and late  snapshots of the Supplementary Movies showing the advection of heavy particles by the viscous flow considered in Fig.~5(b) (main text). 
(a) Movie 1:  $\widetilde{V}_T = 0.6$ and  ${\mbox{St}}/{\mbox{St}}_c = 0.052$, for which attracting points exist and particles from both the closed and the open flow are levitated.
(b) Movie 2:   $\widetilde{V}_T = 0.8$ and ${\mbox{St}}/{\mbox{St}}_c = 0.25$,  scenario similar to (c) for a different parameter choice.
(c) Movie 3:   $\widetilde{V}_T = 0.1$ and ${\mbox{St}}/{\mbox{St}}_c = 1.77$, for which attracting points exist and levitate particles from the closed flow, but no particles from the open flow can approach them.
(d) Movie 4:   $\widetilde{V}_T = 0.6$ and  ${\mbox{St}}/{\mbox{St}}_c = 1.88$, for which no attracting points exist and all particles are centrifuged away from the vortices.
In all panels, blue (red) represent particles from the closed (open) flow, where we choose to show only red in the overlapping areas,  and solid curves mark streamlines of the flow. 
}
       \label{figS2}
\end{figure}

Supplementary Movies 1, 2, 3, and 4 show the particle and fluid dynamics for the scenarios of Navier-Stokes simulations exemplified in Fig.~5(b) (main text).  Each movie corresponds to a different choice of $\widetilde{V}_T$ and ${\mbox{St}}$ for the same flow.  Figure~\ref{figS2} shows  snapshots for each movie, with the corresponding parameters indicated in the caption.

\bigskip
\bigskip
\medskip
{\large\bf Supplementary References}


\begin{thebibliography}{10}
\bibitem{reference_comment}
Because our calculations are for the vortices ($A$, $B$), this conclusion would be reversed for the image vortices ($A'$, $B'$).
\bibitem{comment}
In terms of the particle radius $d$  
and particle-to-fluid density ratio $\rho_p/\rho_f$, for given $\varepsilon$ and flow Reynolds number the Stokes number for heavy particles is
$\St\propto d^2\rho_p/\rho_f$,
 meaning that the predictions can be the same for a denser (but smaller) particle.
\bibitem{comment_aggremment}
The agreement is less satisfactory for counter-rotating vortices (unless $|\gamma |$ is small) because our theory requires not only that  $\vepsi \ll 1$ (i.e., that the distance between $I$ and the wall be large compared to the separation between the vortices) but also  that $\vepsi < (\gamma+1)/2$ (i.e., that each individual vortex be far from the wall---some distance from the wall would be  required also for the stability of the vortex system~\cite{Acheson2000}). The latter condition is satisfied in the co-rotating case since $(\gamma+1)/2> 1/2$, but not in the counter-rotating case unless $| \gamma |$ is sufficiently small, which corresponds to one vortex being sufficiently weaker than the other.
\end{thebibliography}
\end{document}